\documentclass[structabstract]{aa}  
\usepackage{graphicx, natbib, txfonts}
\bibpunct{(}{)}{;}{a}{}{,}
\usepackage{txfonts}

\begin{document}

   \titlerunning{Determining the forsterite abundance in AGB stars}
   \title{Determining the forsterite abundance of the dust around Asymptotic Giant Branch stars}

\author{B.L. de Vries \inst{1, 2}
	\and M. Min \inst{2}
	\and L.B.F.M. Waters \inst{2, 1} 
	\and J.A.D.L. Blommaert \inst{1}
	\and F. Kemper \inst{3}
}

\authorrunning{B.L. de Vries et al.}

\institute{ Instituut voor Sterrenkunde, K.U. Leuven, Celestijnenlaan 200D, 3001 Leuven, Belgium
	\and Sterrenkundig Instituut Anton Pannekoek, University of Amsterdam, Science Park 904, 1098 XH, Amsterdam, The Netherlands
	\and Jodrell Bank Centre for Astrophysics, Alan Turing Building, School of Physics and Astronomy, University of Manchester, Oxford Road, Manchester, M13 9PL, UK}

   \date{Received 3 november, 2009; accepted 10 march 2010}
   
  \abstract
{}
{We present a diagnostic tool to determine the abundance of the crystalline silicate forsterite in AGB stars surrounded by a thick shell of silicate dust. Using six infrared spectra of high mass-loss oxygen rich AGB stars we obtain the forsterite abundance of their dust shells.}
{We use a monte carlo radiative transfer code to calculate infrared spectra of dust enshrouded AGB stars. We vary the dust composition, mass-loss rate and outer radius. We focus on the strength of the 11.3 and the 33.6 $\mu$m forsterite bands, that probe the  most recent (11.3 $\mu$m) and older (33.6 $\mu$m) mass-loss history of the star. Simple diagnostic diagrams are derived, allowing direct comparison to observed band strengths.}
{Our analysis shows that the 11.3 $\mu$m forsterite band is a robust indicator for the forsterite abundance of the current mass-loss period for AGB stars with an optically thick dust shell. The 33.6 $\mu$m band of forsterite is sensitive to changes in the density and the geometry of the emitting dust shell, and so a less robust indicator. Applying our method to six high mass-loss rate AGB stars shows that AGB stars can have forsterite abundances of 12\% by mass and higher, which is more than the previously found maximum abundance of 5\%.}
   {}


   \maketitle

\section{Introduction} 

Asymptotic Giant Branch (AGB) stars represent a late phase in the evolution of low- and intermediate mass stars (0.8-8.0\,M$_{\odot}$). At the end of the Red Giant Branch phase when the helium in the core is depleted a star enters the AGB phase, with a degenerate carbon and oxygen core surrounded by a helium and a hydrogen burning shell. The burning shells are surrounded by a convective envelope and a dynamically active atmosphere \citep{HO03}.
Depending on the carbon-to-oxygen ratio, either oxygen or carbon is locked in the CO molecule. This leads to two possible chemistries. Initially AGB stars exhibit an oxygen-rich chemistry reflecting the original composition of the natal molecular cloud. An AGB star becomes C-rich when sufficient carbon is dredged up from the helium-burning shell where it is formed by nucleosynthesis processes. In this study we will only consider O-rich AGB stars.

One of the most important aspects of AGB stars is their stellar wind. AGB stars can have mass-loss rates between $10^{-8}$ \nolinebreak M$_{\odot}$ \nolinebreak yr$^{-1}$ and about $10^{-4}$ M$_{\odot}$ yr$^{-1}$. The high mass-loss rate automatically constrains the life-time of these objects, and gives rise to dense circumstellar envelopes. Two mechanisms contribute to this mass-loss. First matter gets accelerated outwards due to shocks \citep{HO03}. When the material reaches distances where the temperature is low enough the material can condense into solid-state particles (from here on called dust). Then a second acceleration mechanism occurs: radiation pressure on the dust grains. The gas is coupled to the dust and dragged along. This leads to expansion speeds of 10 km s$^{-1}$ and higher.

Even though amorphous silicate dust is more abundant, crystalline silicate dust is also seen in many astronomical environments like disks around pre-main-sequence stars \citep{waelkens96, meeus01,spitzer1}, comets \citep{wooden02}, post-main-sequence stars \citep{waters96, syl99, mol02} and active galaxies \citep{kemper07, spoon06}. Theoretical studies predict that the formation of crystalline silicate dust is dependent on the gas density \citep{tielens98,gailsedl99, sogawa99}. Higher densities are more favorable for the formation of crystals than low densities. \citet{speck08} also argue that there is a correlation between crystallinity, mass-loss rate and the initial mass of the star.

These suggestions are consistent with the detection of crystalline silicates in the high density winds of OH/IR stars and the disks around post-AGB stars. Indeed, observations show a lack of crystalline silicate dust bands in the spectra of low mass-loss rate AGB stars \citep{waters96, syl99}.
However, an alternative explanation for this lack exists. \citet{kemper01} have shown that a contrast effect hinders the detection of crystalline silicate bands in infrared spectra of low mass-loss rate AGB stars. This contrast effect is further discussed in sect. 3.2.2.

\citet{kemper01} obtained an indication for the crystalline silicate abundance in the outflow of AGB stars by comparing the strength of crystalline silicate emission bands of observations to those of model spectra. For the crystalline silicates enstatite and forsterite they measured a maximum abundance of $\sim$5\% by mass. 

AGB stars are the main contributors of crystalline silicate material to the ISM. The creation of crystalline silicates in the ISM is not possible since the temperatures are too low. Therefore the hypothesis would be that the abundance of crystalline silicate dust in the ISM would be comparable to that found in the dust shells of AGB stars. The crystalline silicate abundance in the diffuse ISM has been investigated by \citet{kemper04, kemperErr05}. They looked at the line of sight towards the Galactic Centre and found an upper limit of the degree of crystallinity in the diffuse ISM of 2.2\% by mass. The abundance found in AGB stars by \citet{kemper01} is therefore higher than that for the diffuse ISM, suggesting the crystalline material may be effectively destroyed or hidden. Possible mechanisms could be amorphization by inclusion of iron atoms in the crystalline lattice, shocks and UV radiation (see \citet{kemper04, kemperErr05} for a discussion on such mechanisms).

Determining the crystalline fraction of silicates in AGB stars, as a function of mass-loss rate, will enhance our understanding in two areas: First, what are the conditions needed for the formation of crystalline silicates? Dust condensation theory suggests that the density is critical in the formation of crystalline silicates \citep{tielens98,gailsedl99, sogawa99}, but this has still not been verified by observations. Secondly, how much crystalline silicate material is deposited by AGB stars to the diffuse ISM? Do the abundances in the outflow of AGB stars and the ISM match?

The most thorough way to determine the forsterite abundance in the outflows of AGB stars is to fit the full spectral range offered by ISO or Spitzer spectra. This is an exhausting endeavor to apply to multiple stars and it does not necessary deliver a good understanding on how the spectral bands depend on dust shell parameters. Therefore it is useful to follow and expand the approach of \citet{kemper01} who derived diagnostics for the 33.6 $\mu$m forsterite band. We expand on this study by carefully studying the dependence of the spectral bands on the mass-loss rate, dust composition, thermal contact between grains and we also use two well-chosen bands, one probing the most recent mass-loss (11.3 $\mu$m) and one probing the older, colder mass-loss (33.6 $\mu$m). We derive a diagnostic diagram that is fairly easy to use and allows one to derive the forsterite abundance in the AGB star outflows. 

In sect. 2 we outline how the models were computed and show several examples of computed spectra. The strength determination of the two forsterite bands is discussed in sect. 3. Also the diagnostic diagram to measure the forsterite abundance is introduced and possible limitations are discussed. In sect. 4 we apply the diagnostic diagram to six ISO-SWS spectra of AGB stars.

\section{Modeling the AGB dust shell} 

In this section we present the setup used to compute model spectra and show a few examples of infrared spectra.

The model spectra were computed using the Monte Carlo radiative transport code MCMax \citep{min09}. Within the code we treat AGB stars as a central star enclosed in a spherically symmetric dust shell. The central star is defined by its radius and spectrum. A spectrum of an M9III giant (2667K) is used for the central star \citep{fluks94}. For the luminosity a typical value of 7000 $\rm{L}_{\odot}$ is used \citep{HO03}. Using this luminosity and integrating the spectrum one obtains a radius of $395\,\rm{R}_{\odot}$ for the central star.

An $r^{-2}$ density distribution for the dust shell gives rise to a time-independent outflow at a constant velocity. There is ample evidence for strong time variability of AGB mass-loss, see for example \citet{decin07}, which would correspond to a different density structure than the one adopted here. We will come back to this point in section 3.2.3. 

The terminal wind speed (v$_{\mathrm{exp}}$) is set to the typical value of $10$ km s$^{-1}$. For the dust-to-gas ratio a value of 1/100 is assumed. For our models the dust-to-gas ratio only determines the amount of material in the dust shell that is in the form of dust. Changing this value will result in a different dust mass-loss rate and therefore it only scales the total mass-loss rates used in this paper. The temperature of the dust shell reaches a value of 1000 K at a radius of $\sim$10 AU. This radius is taken to be the radius at which the amorphous silicate condenses, and therefore defines the inner radius of the dust shell. Doing this we assume that the dust species formed before this 1000 K limit have a negligible effect on the spectrum. This is especially true for the optically thick dust shells where the infrared spectrum is not sensitive anymore to changes in the inner part of the dust shell. We note that in optically thin dust shells the maximum dust temperature detected in a study by  \citet{heras05} is around 700 K. So the calculations take somewhat high condensation temperatures for optically thin dust shells.

\subsection{Dust opacities}

In this paper we only consider the major components of the dust in AGB outflows. We use silicate grains, both in crystalline and amorphous form and metallic iron grains. To calculate the opacities of dust grains, the chemical composition, lattice structure and shape distribution have to be chosen. These choices will be discussed in this section.

The composition of the amorphous silicate component in AGB stars is currently not well known. In this paper we fix the amorphous silicate composition to be consistent with the spectroscopic signature of the low mass-loss rate AGB star Mira. Mira shows no clear signs of crystalline silicates making it an ideal case to study the amorphous component in AGB outflows. We have fitted the ISO-SWS infrared spectrum of Mira using amorphous silicates with different compositions and find that the best fit was made with 65.7\% MgFeSiO$_{4}$ and 34.3\% Mg$_{2}$SiO$_{4}$ implying an average composition of Mg$_{1.36}$Fe$_{0.64}$SiO$_{4}$. The fit was made with CDE shaped particles (see below). By adopting the composition for the amorphous silicates as found for Mira we make the implicit assumption that the composition of the amorphous silicate dust grains is not dependent on the mass-loss rate of the AGB star.

\begin{figure}
		\resizebox{\hsize}{!}{\includegraphics{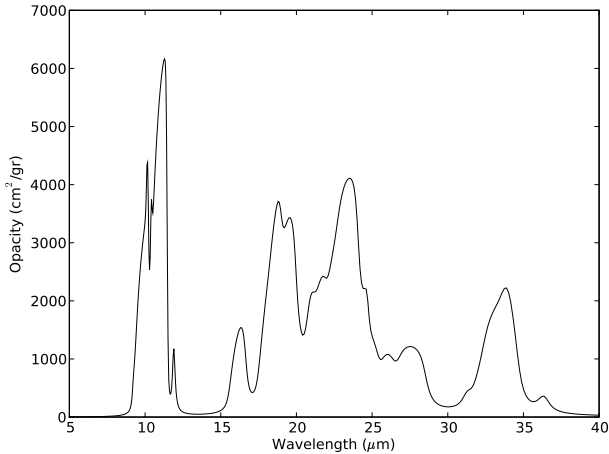}}
		\caption{The opacities of the crystalline silicate forsterite. They clearly show the strong bands at 11.3 $\mu$m and 33.6 $\mu$m bands which are investigated in this study as possible forsterite abundance indicators. The opacities are calculated using the complex refractive indices taken from \citet{servoin73} and using CDE particle shapes.}
		\label{kappas}
		\end{figure}

For the refractive indices we have used data measured in the laboratory for a large number of material compositions. We use here for the amorphous silicates the data from \citet{jager2003} and \citet{dorschner95}, the forsterite data from \citet{servoin73}, the enstatite data from \citet{jager98} and the metallic iron laboratory data from \citet{ordal88}. We always use an equal abundance of enstatite and forsterite. The opacities of forsterite are shown in fig. \ref{kappas}.

\citet{min03} have shown that the use of homogeneous spheres as particle shapes introduces unrealistic effects. This is because the symmetries in these particles introduce resonances. It has been shown by several authors that the CDE (Continuous Distribution of Ellipsoids, \citet{BH83}) shape distribution provides a very good fit to both astronomical observations and laboratory measurements. Therefore, in this study we use the CDE shape distribution.

Laboratory experiments show that crystalline silicates,  metallic iron and amorphous silicates have different condensation temperatures, which are also pressure dependent  \citep{gailsedl99}. It is very well possible that high temperature condensates may act as seeds for lower temperature materials to condense onto \citep{gailsedl99}. Therefore we must consider chemically inhomogeneous grains as a possibility, since this will affect their temperatures and so their IR emission. However, grain condensation is likely to be a dynamical process which is difficult to model, and so it is hard to predict from chemical equilibrium condensation theory what the nature of the chemically inhomogeneous grains may be. 
Given these uncertainties we decided to use two choices for chemically inhomogeneous grains and investigate the effects of these two choices on the calculated spectra. These two choices should span most of parameter space. Our standard model assumes that amorphous silicate grains condense first, followed by metallic iron that forms on top of the amorphous silicates, along the lines suggested by \citet{gailsedl99}. Therefore metallic iron and amorphous silicates are in thermal contact and have the same temperature in our calculations. We assume the crystalline silicates to remain isolated and so not in thermal contact with other dust species. This minimizes the temperature of these grains, due to the low opacity at near-IR and optical wavelengths. Our second choice is to assume that all grains are in thermal contact and so have the same temperature. This obviously eliminates any temperature differences and probes the extreme case in which crystalline silicates, despite their low opacities at short wavelengths, are still efficiently heated by stellar photons. 


\subsection{Model spectra} 

In this section we show examples of computed spectra. This will illustrate the effect of the mass-loss rate, metallic iron abundance, forsterite abundance and the outer radius of the dust shell on the infrared spectra. A standard set of model parameters is used (see table \ref{table:1}) and for the rest of this paper, if the parameters are not mentioned explicitly, they are set to these standard values.

For the metallic iron abundance a standard value of 4\% is chosen. This is consistent with \citet{kemper02}, who have shown that the infrared spectrum of OH127.8+0.0 can be explained by including that amount of metallic iron in their models. A starting value for the forsterite and enstatite abundance has been chosen to be 4\%. In most studies the outer radius of the dust shell is practically set to infinity. In this study we want to treat the outer radius as a variable, in order to investigate its effect. As a standard value we have chosen an outer radius of 500 AU. An interesting example is that \citet{ches05} have shown that the dust shell of the current mass-loss of OH 26.5+0.6 is about 250 yr old. Using a 10 km s$^{-1}$ outflow velocity this translates to an outer radius of the dust shell of roughly 500 AU.

		\begin{table*}
		\caption{List of model parameters and their standard values and the range over which the parameters are varied.} 
		\label{table:1} 
		\centering 
		\begin{tabular}{l l l l} 
		\hline\hline 
		Parameter			& 	Standard value				& Minimum	& Maximum	\\
		\hline 

		Mass-loss			&	$3 \cdot 10^{-5} M_{\odot} $yr$^{-1}$	& $1 \cdot 10^{-6} M_{\odot} $yr$^{-1}$	& $1 \cdot 10^{-4} M_{\odot} $yr$^{-1}$\\
		Metallic iron abundance	&	4\%						& 0\%		& 9\%		\\
		Forsterite abundance		&	4\%						& 0\%		& 10\%		\\
		Enstatite abundance		&	set equal to forsterite			& 0\%		& 10\%		\\
		Outer radius			&	500 AU						& 300 AU	& 3000 AU	\\

		\hline 
		\end{tabular}
		\end{table*}

\begin{figure}
		\resizebox{\hsize}{!}{\includegraphics{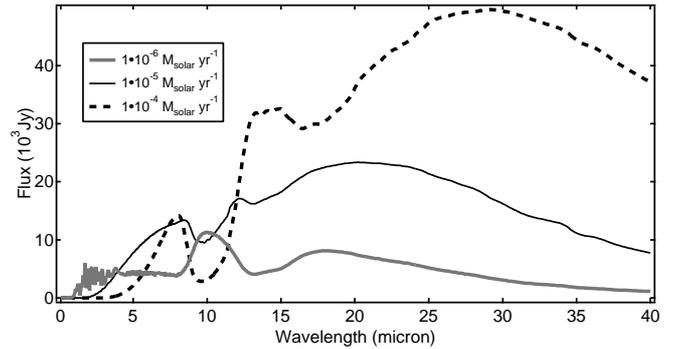}}
		\caption{Three model spectra with different mass-loss rates are shown. The solid grey, solid black and dashed black line have mass-loss rates of $1 \cdot 10^{-6} \,\rm{M}_{\odot}\,$yr$^{-1}$, $1 \cdot 10^{-5} \,\rm{M}_{\odot}\,$yr$^{-1}$ and $1 \cdot 10^{-4} \,\rm{M}_{\odot}\,$yr$^{-1}$ respectively. All model parameters other than the mass-loss rate are set to standard values (see table \ref{table:1}).}
		\label{mdot_compressed}
		\end{figure}

Figure \ref{mdot_compressed} shows three spectra with different mass-loss rates. These mass-loss rates are chosen to represent objects with the 9.7 $\mu$m band in emission, partially in absorption and in absorption. The spectrum with a mass-loss rate of $10^{-6}$ M$_{\odot}$ yr$^{-1}$ shows both the 9.7 $\mu$m and the 18 $\mu$m band of amorphous silicate in emission. In this case the dust shell is optically thin and the spectrum of the central star can be observed. Increasing the mass-loss rate to $10^{-5}$ M$_{\odot}$ yr$^{-1}$ makes the blue part of the spectrum optically thick, obscuring the spectrum of the central star and causing the 9.7 $\mu$m band to go into self-absorption. The spectrum of the $10^{-4}$ M$_{\odot}$ yr$^{-1}$ case has the 9.7 $\mu$m and the 18 $\mu$m band of amorphous silicate firmly in absorption.

\begin{figure}
		\resizebox{\hsize}{!}{\includegraphics{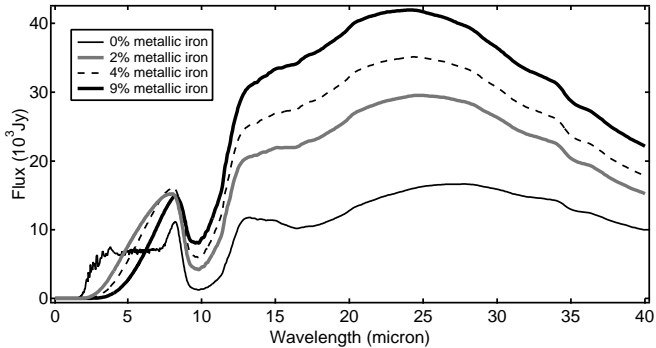}}
		\caption{Four spectra with different metallic iron abundances are shown. The solid black, solid grey, dashed black and bold black lines have metallic iron abundances of 0\%, 2\%, 4\% and 9\% respectively. All model parameters other than the metallic iron abundance rate are set to standard values (see table \ref{table:1}).}
		\label{ironCDE}
		\end{figure}

Spectra with different metallic iron abundances are shown in fig. \ref{ironCDE}. Metallic iron absorbs short wavelength radiation more efficiently and since these grains are in thermal contact with the amorphous silicates, both heat up. In turn the amorphous silicates radiate this energy at longer wavelengths. It can be seen from the spectra that metallic iron influences the opacity in the 3 to 8 $\mu$m region and that the emission at longer wavelengths is increased when metallic iron is added.

\begin{figure}
		\resizebox{\hsize}{!}{\includegraphics{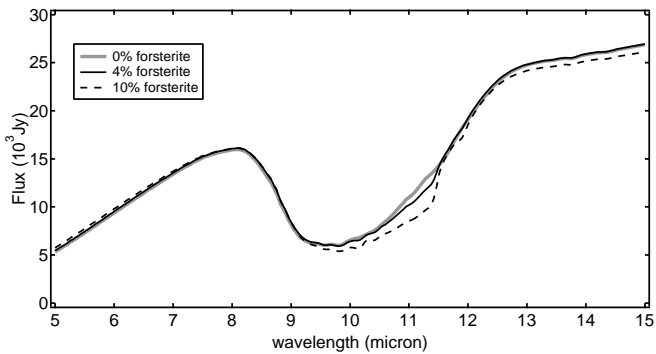}}
		\caption{The 9.7 $\mu$m band of amorphous silicate of spectra with different forsterite abundances are shown. The solid grey, solid black and dashed black lines have forsterite abundances of 0\%, 4\% and 10\% respectively. All model parameters other than the forsterite abundance are set to standard values (see table \ref{table:1}).}
		\label{fo_10zoom}
		\end{figure}

Figure \ref{fo_10zoom} shows model results for different forsterite abundances. The forsterite band at 11.3 $\mu$m can clearly be recognized as a shoulder on the red side of the 9.7 $\mu$m band of amorphous silicate, and increases in strength with increasing abundance.

\begin{figure}
		\resizebox{\hsize}{!}{\includegraphics{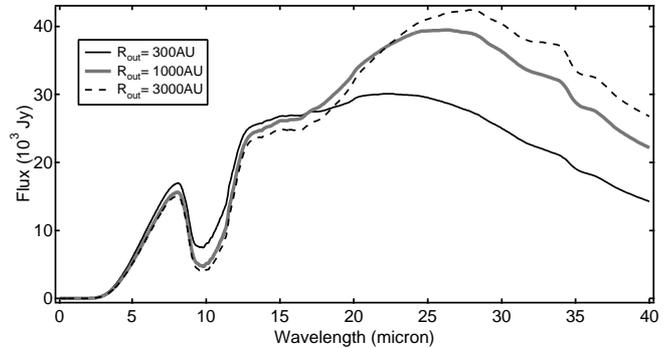}}
		\caption{The effect of the outer radius of the dust shell on the spectra is plotted. The solid black, solid grey and dashed black lines have outer radii of 300AU, 1000AU and 3000AU respectively. All model parameters other than the outer radius are set to standard values (see table \ref{table:1}).}
		\label{Rout}
		\end{figure}

To investigate the influence of the outer radius on the spectrum, models were run with outer radii up to 3000 AU (figure \ref{Rout}). From the figure it can be seen that changing the outer radius has a significant impact. By increasing the outer radius there will be more cold dust and therefore the spectrum becomes redder. The outer radius also has a profound influence on the strength of the 33.6 $\mu$m band. For an outer radius of 3000AU the band is seen to be strong and easily detectable. If the radius is decreased to 300AU the strength of the band is also seen to go down steeply. We will see this effect more quantitatively in sect. 3.2.2.

\section{Forsterite abundance indicators}

In this section we discuss the applicability of the 11.3 $\mu$m and the 33.6 $\mu$m forsterite bands as forsterite abundance indicators. We compute the strength of these bands for model spectra and study their dependence on the model parameters. For both bands the strength is measured in spectra with varying mass-loss rate, forsterite abundance, metallic iron abundance and outer radius. We have chosen to measure the equivalent width as a measure of the strength of the band because it is less sensitive to a choice in grain shapes, than for instance the ratio of the peak flux over the continuum. Although the band shape may change for different grain shapes, the total power remains conserved.

Both bands require a different method for measuring their strength. The 11.3 $\mu$m band is an absorption band superimposed on the 9.7 $\mu$m band of amorphous silicate and a careful measurement of its strength is needed. In our method to measure the strength of the 11.3 $\mu$m band we limit this study to the case where the 9.7 $\mu$m band of amorphous silicate is in absorption. As a measure of the strength of the band, the equivalent width is calculated from the optical depth profile (see sect. 3.1).

The 33.6 $\mu$m band is an emission band for the mass-loss rates that are considered. This means the band is predominantly formed in the optically thin region of the wind. Constructing a continuum under the band enables us to calculate the equivalent width. The details on how this is calculated are discussed in sect. 3.2.

Comparable to the 11.3 $\mu$m band of forsterite, enstatite has a band at 9.3 $\mu$m. We are looking into the applicability of this enstatite band, but its analysis is more complicated than for the 11.3 $\mu$m band of forsterite. The 9.3 $\mu$m band of enstatite lies closer to the absorption maximum of the 9.7 $\mu$m band of amorphous silicate. In this paper we focus on the bands of forsterite only, since these can be compared with the results of \citet{kemper01}.

\subsection{The 11.3 $\mu$m forsterite band}
\subsubsection{Method}
To measure the strength of the 11.3 $\mu$m band a local continuum is defined on top of the 9.7 $\mu$m amorphous silicate band. A linear interpolation between the flux densities at 8.0 $\mu $m to 8.1 $\mu$m and 12.9 $\mu$m to 13.0 $\mu$m is made to construct this local continuum. 

\begin{figure}
		\resizebox{\hsize}{!}{\includegraphics{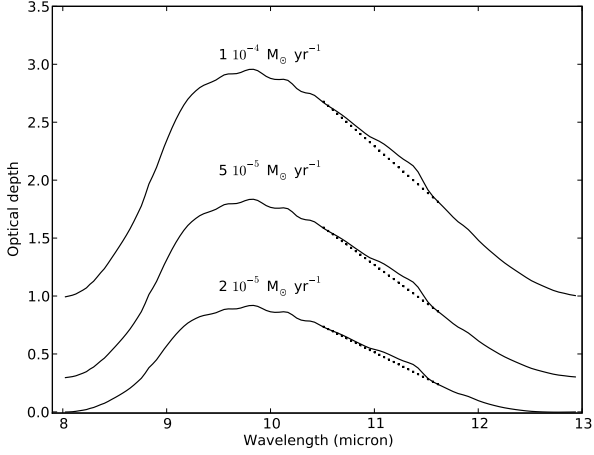}}
			\caption{Three optical depth curves of the 9.7 $\mu$m band of amorphous silicate for three different mass-loss rates. The constructed local ``continuum'' in order to extract the 11.3 $\mu$m band is also shown. The curves are spaced vertically for clarity. The bottom, middle and top curve have mass-loss rates of $2 \cdot 10^{-5} \,\rm{M}_{\odot}\,$yr$^{-1}$, $5 \cdot 10^{-5} \,\rm{M}_{\odot}\,$yr$^{-1}$ and $1 \cdot 10^{-4} \,\rm{M}_{\odot}\,$yr$^{-1}$, respectively. All model parameters not mentioned in the graph are set to standard values (see table \ref{table:1}) }
		\label{97tau}
		\end{figure}

\begin{figure}
		\resizebox{\hsize}{!}{\includegraphics{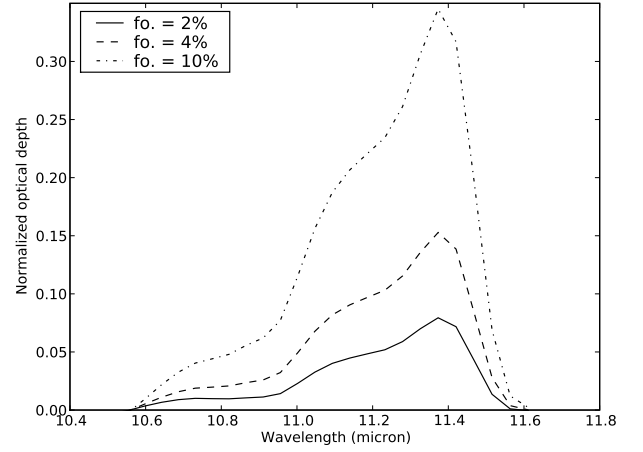}}
			\caption{Three normalized optical depth curves of the 11.3 $\mu$m band of forsterite for three different forsterite abundances. From the solid, dashed, dash-dotted curves have a forsterite abundance of 2\%, 4\% and 10\% respectively. All model parameters not mentioned in the graph are set to standard values (see table \ref{table:1}) }
		\label{113tau}
		\end{figure}

For every wavelength in the range of the 9.7 $\mu$m band the optical depth is calculated: $F(\lambda)/F_{cont}(\lambda) = e^{-\tau}$, where $F(\lambda)$ is the flux of the model and $F_{cont}(\lambda)$ the flux of the continuum. In the optical depth profile $\tau(\lambda)$ a local ``continuum'' is defined under the 11.3 $\mu$m band. This is done by making a linear fit through the optical depths at the wavelength points 10.5 $\mu$m to 10.6 $\mu$m and 11.6 $\mu$m to 11.7 $\mu$m. Figure \ref{97tau} shows three optical depth curves together with the continuum constructed under the 11.3 $\mu$m band. The 11.3 $\mu$m band can now be normalized: $\tau(\lambda)/\tau_{cont}(\lambda) -1$. The equivalent width in terms of optical depth is calculated by numerical integration from 10.5 $\mu$m to 11.7 $\mu$m. Figure \ref{113tau} shows three normalized optical depth curves for the 11.3 $\mu$m band.

		\begin{figure}
		\resizebox{\hsize}{!}{\includegraphics{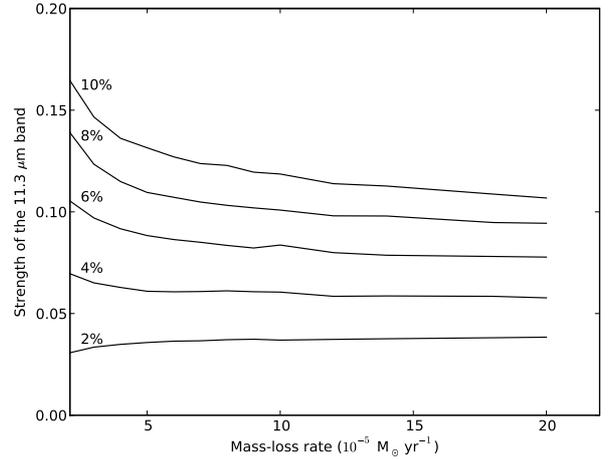}}
		\caption{The equivalent width of the 11.3 $\mu$m band of forsterite as a function of the mass-loss rate. The different curves have different forsterite abundances. All model parameters not mentioned in the graph are set to standard values (see table \ref{table:1}). The upward trend seen for the 2\% forsterite curve in the mass-loss rate range of $1 \cdot 10^{-5} \,\rm{M}_{\odot}\,$yr$^{-1}$ to $5 \cdot 10^{-5} \,\rm{M}_{\odot}\,$yr$^{-1}$ is due to the approximate way in which the continuum is estimated in order to calculate the strength of the 11.3 $\mu$m  (see section 3.1.2). }
		\label{113_cryst_th}
		\end{figure}

\subsubsection{Results}

The effect of changes in the mass-loss rate on the strength of the 11.3 $\mu$m band is shown in fig. \ref{113_cryst_th}, for a range of forsterite abundances. The band strength decreases with increasing mass-loss, reaching a constant value for the highest mass-loss rates. This behavior can be understood as follows. For the highest mass-loss rates, optical depths are high for many lines of sight through the dust shell. In this case the band strength is determined by the ratios of the optical depths of the dust species, and is not sensitive any more to temperature differences. This is why the curves in fig. \ref{113_cryst_th} reach a constant value. At lower mass-loss rates, \emph{emission}  from optically thin lines of sight through the dust shell become more important. The temperature difference between dust species plays a role and the warmer amorphous silicates and metallic iron will begin to fill in the 9.7 $\mu$m absorption band seen so prominently at high mass-loss rates. This results in a relatively smaller optical depth of the amorphous silicate band compared to that of the crystalline silicates, and so the ratio ($\tau (\lambda) / \tau_{cont}(\lambda) - 1$) will increase. The curve for 2\% forsterite does not follow the general trend seen in fig. \ref{113_cryst_th} at low mass-loss rates. This is due to the approximate way in which the continuum is estimated (linear fit), causing small errors that can be important for weak bands. 

The outer radius of the dust shell has no significant effect on the strength of the 11.3 $\mu$m band and a figure has been omitted.

\begin{figure}
		\resizebox{\hsize}{!}{\includegraphics{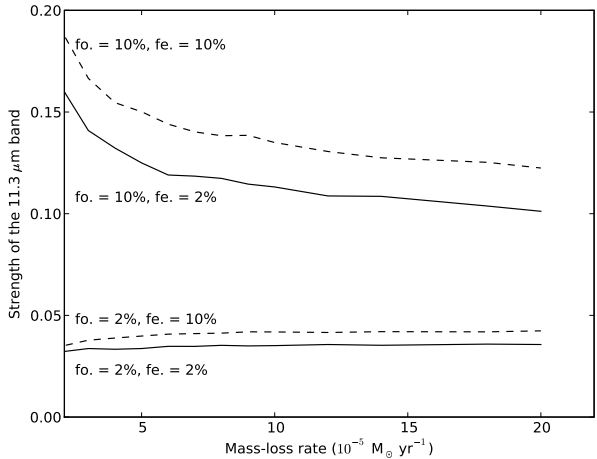}}
		\caption{The effect of changes in the metallic iron abundance on the strength of the 11.3 $\mu$m forsterite band for two choices of the forsterite abundance of 2\% and 10\% respectively. Solid lines represent 2\% metallic iron, dashed lines 10\%. All model parameters not mentioned in the graph are set to standard values (see table \ref{table:1}).}
		\label{113_iron_th}
		\end{figure}

Adding metallic iron increases the emission of the continuum under the 9.7 $\mu$m band. This reduces the optical depth in the region of the 9.7 $\mu$m band ($\tau_{cont}(\lambda)$), causing the equivalent width of the 11.3 $\mu$m band to increase (see fig. \ref{113_iron_th}).

\begin{figure}
		\resizebox{\hsize}{!}{\includegraphics{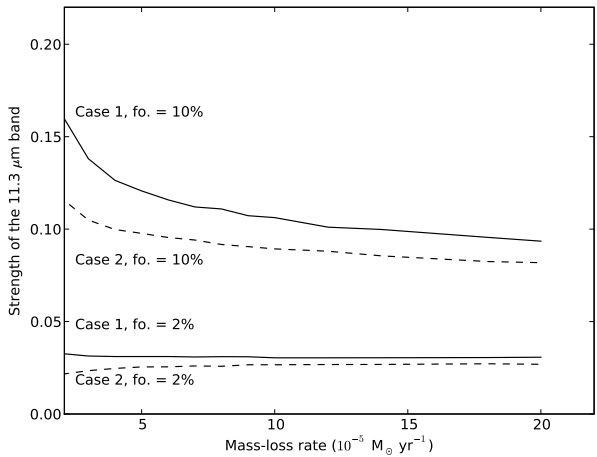}}
		\caption{The equivalent width of the 11.3 $\mu$m band is plotted as a function of the mass-loss rate. It shows the 2\% and the 10\% forsterite abundance curves for two different cases. The first case is where only metallic iron and amorphous silicate grains are in thermal contact (solid line). In the second case all dust species are in thermal contact (dashed line). All model parameters not mentioned in the graph are set to standard values (see table \ref{table:1}).}
		\label{113_th_cont}
		\end{figure}

In order to test the effects of our choice regarding thermal contact we also consider the case of thermal contact between all grains (section 2.1). Figure \ref{113_th_cont} compares the strength of the 11.3 $\mu$m band in the case where only amorphous silicate and metallic iron are in thermal contact with the case where all dust species are in thermal contact. The strength of the 11.3 $\mu$m band decreases by adopting thermal contact between all the dust species. This is because part of the thermal energy is redistributed to the crystalline silicate species, which do not contribute to the continuum. This causes the flux of the continuum to drop, resulting in a weaker 11.3 $\mu$m band.

\subsubsection{Discussion}

The 11.3 $\mu$m forsterite band is formed close to the star and so it probes the most recent mass-loss rate. We find that its strength is relatively insensitive to model parameters (other than the forsterite abundance itself)  when the mass-loss rate exceeds roughly $3 \cdot 10^{-5} \,\rm{M}_{\odot}\,$yr$^{-1}$. The largest uncertainty at these high mass-loss rates is due to changes in the metallic iron abundance. The effect of thermal contact between forsterite and other dust species is minor (1-2\%). In our standard model we assume no thermal contact, which leads to conservative estimates of the forsterite abundance.  

\subsection{The 33.6 $\mu$m forsterite band}
\subsubsection{Method}

\begin{figure}
		\resizebox{\hsize}{!}{\includegraphics[width=60mm]{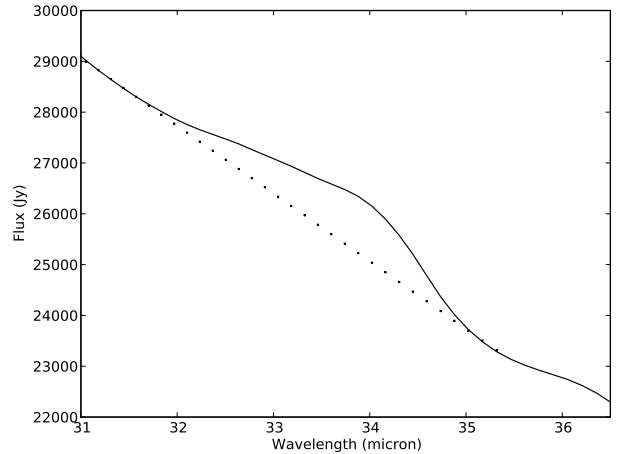}}
				\caption{The continuum constructed under the 33.6 $\mu$m band of forsterite for a model spectrum with standard parameters (see table \ref{table:1})}
				\label{336_cont}
		\end{figure}

\begin{figure}
		\resizebox{\hsize}{!}{\includegraphics[width=60mm]{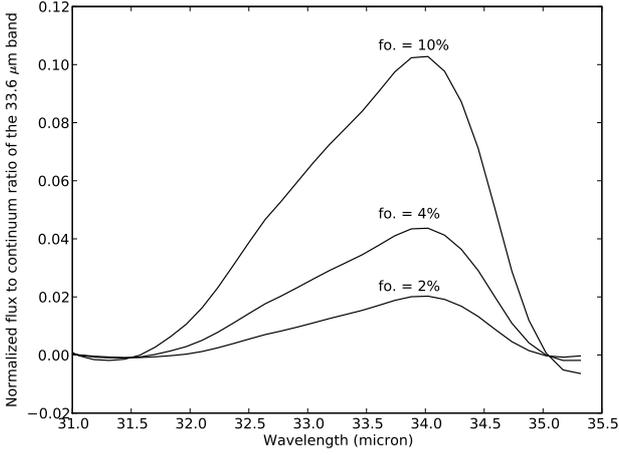}}
				\caption{Normalized 33.6 $\mu$m bands for different forsterite (fo.) abundances. Other parameters are set to standard values (see table \ref{table:1})}
				\label{336_band}
		\end{figure}

In order to measure the equivalent width of the 33.6 $\mu$m band a local continuum is defined by making a linear fit through the flux densities at 31.0 $\mu$m to 31.5 $\mu$m and 35 $\mu$m to 35.5 $\mu$m. An example of such a continuum is shown in fig. \ref{336_cont}. This continuum is used to normalize the spectrum in this region, calculating $F(\lambda)/F_{cont}(\lambda) -1$. Three such normalized bands with different forsterite abundances are shown in fig. \ref{336_band}. The equivalent width of the normalized band is calculated by numerical integration between the wavelength values of 31 $\mu$m and 35.5 $\mu$m. We note that due to the emission nature of the band its equivalent width will be sensitive to the abundance and temperature of the dust species emitting at 33.6 $\mu$m.

		\begin{figure}
		\resizebox{\hsize}{!}{\includegraphics[width=60mm]{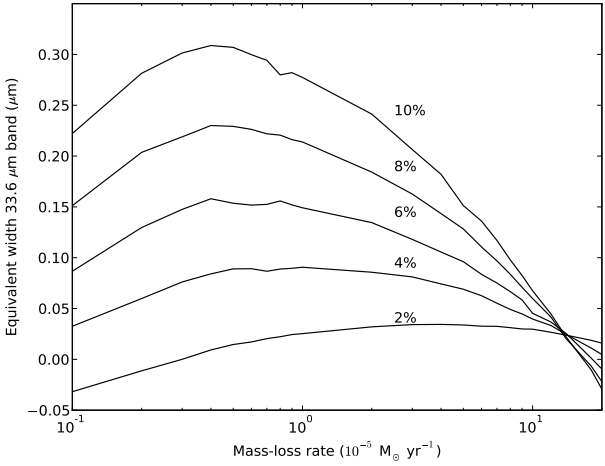}}
				\caption{The equivalent width of the 33.6 $\mu$m band of forsterite as a function of the mass-loss rate. The different curves have different forsterite abundances. All model parameters not mentioned in the graph are set to standard values (see table \ref{table:1}).}
				\label{336_cryst_th}
		\end{figure}

\subsubsection{Results}

Figure \ref{336_cryst_th} shows the equivalent width of the 33.6 $\mu$m band as function of mass-loss rate for different forsterite abundances. The mass-loss rates range from $10^{-6} \,\rm{M}_{\odot}\,$yr$^{-1}$ up to $2 \cdot 10^{-4} \,\rm{M}_{\odot}\,$yr$^{-1}$. The strength of the band initially increases with increasing mass-loss rate, followed by a drop and eventually the band is in absorption. This behavior is similar to that seen by \citet{kemper01} and has been explained by these authors in terms of changes in the temperature of the different dust species as a function of mass-loss rate. For low mass-loss rates, the dust shell is optically thin and the (near-IR) stellar photons heat the dust directly. This causes large temperature differences between amorphous silicate (which has high opacity at near-IR wavelengths) and crystalline silicates (which are almost transparent in the near-IR). The band strength of forsterite will thus be weak. For higher mass-loss rates the dust shell will become more optically thick and more and more of the dust will receive longer wavelength photons emitted by the optically thick inner dust shell. The opacity of amorphous and crystalline dust at mid-IR wavelengths is very comparable and so the temperature difference between them will decrease. This results in an increase of the band strength. For even higher mass loss rates (larger than about $10^{-5} \,\rm{M}_{\odot}\,$yr$^{-1}$) some lines of sight become optically thick at 33.6 $\mu$m and this will result in a weakening of the band, eventually even leading to absorption. 

The fact that the curve for 2\% forsterite becomes negative at low mass-loss rates is due to the construction of the continuum, which is curved but approximated by a linear fit. This introduces small errors that can be important for weak bands, especially at low mass-loss rates.

		\begin{figure}
		\resizebox{\hsize}{!}{\includegraphics{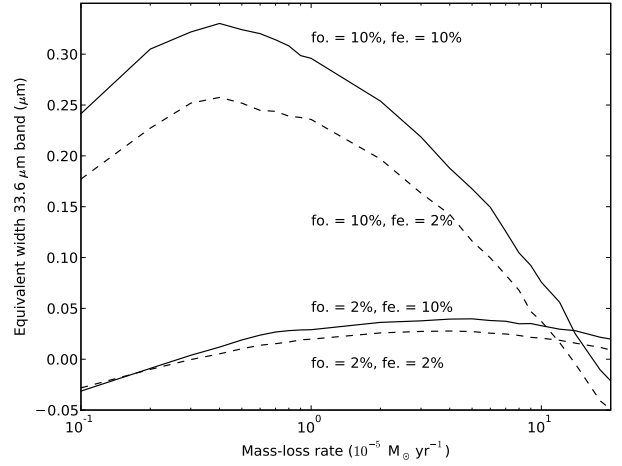}}
		\caption{ The equivalent width of the 33.6 $\mu$m band as a function of the mass-loss. Both the curves with 2\% and 10\% forsterite are shown for the metallic iron abundances of 2\% (dashed line) and 10\% (solid line). All model parameters not mentioned in the graph are set to standard values (see table \ref{table:1}).}
		\label{336_iron_th}
		\end{figure}

The effect of metallic iron on the equivalent width of the 33.6 $\mu$m band is demonstrated in fig. \ref{336_iron_th}. Metallic iron increases the flux of the continuum, but not that of forsterite. Since the equivalent width is a measure of the relative strength of forsterite to the continuum, increasing the metallic iron abundance will decrease the equivalent width. The reason for this is that forsterite is considered not to be in thermal contact with the amorphous silicates, but metallic iron is.

		\begin{figure}
		\resizebox{\hsize}{!}{\includegraphics{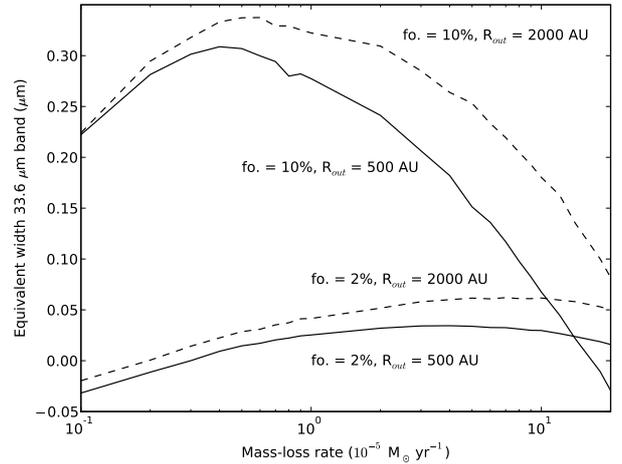}}
			\caption{The equivalent width of the 33.6 $\mu$m band as a function of the mass-loss rate. It shows the 2\% and the 10\% forsterite abundance curves with outer radii of 500 (solid line) and 2000AU (dashed line). All model parameters not mentioned in the graph are set to standard values (see table \ref{table:1}).}
		\label{336_rout_th}
		\end{figure}

Increasing the outer radius of the dust shell increases the amount of cold, emitting dust and therefore increases the emission of the 33.6 $\mu$m band (figure \ref{336_rout_th}). It is interesting that the effect of the size of the dust shell becomes smaller and even negligible for low mass-loss rates. Below $\sim 6\cdot 10^{-6} \,\rm{M}_{\odot}\,$yr$^{-1}$ the effect is already very small. The reason is that the wind is so optically thin that adding more material by increasing the outer radius adds material of such low temperatures that its emission is negligible.

\begin{figure}
		\resizebox{\hsize}{!}{\includegraphics{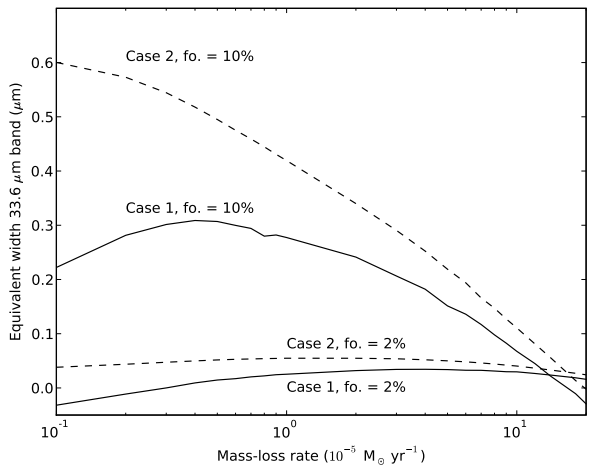}}
			\caption{The equivalent width of the 33.6 $\mu$m band is plotted as a function of the mass-loss rate. It shows the 2\% and the 10\% forsterite abundance curves for two different cases. The first case (solid line) represents models in which only the metallic iron and amorphous silicate grains are in thermal contact. In the second case (dashed line) all the dust species are in thermal contact with each other. All model parameters not mentioned in the graph are set to standard values (see table \ref{table:1}).}
		\label{336_th_cont}
		\end{figure}

Figure \ref{336_th_cont} shows the equivalent width of the 33.6 $\mu$m band for the two different cases of thermal contact (see sect. 3.1.2) for models with 2\% and 10\% forsterite. It shows that the equivalent width of the band goes up when all the dust species are in thermal contact. In the case where all dust species are in thermal contact with each other, forsterite is heated up by metallic iron, creating a stronger emission band. The strength of this effect depends on the forsterite abundance.

\subsubsection{Discussion}

The 33.6 $\mu$m band is sensitive to the mass-loss rate of the AGB star as well the outer radius of the dust shell. The size of the dust shell has an effect because the 33.6 $\mu$m band is mainly formed by the emission of cold dust in the outskirts of the wind. This tells us that other parameters that influence this cold dust will also affect the 33.6 $\mu$m band. As an example we know that AGB stars can have periods with different mass-loss rates and material of such a previous mass-loss period could also influence the 33.6 $\mu$m band. Such time variable mass-loss can be modeled as a sudden jump in the density, but can also be studied by introducing a steeper or flatter radial density gradient. Changes in the radial density gradient will affect the relative strength of the 33.6 $\mu$m emission band with respect to that of the 11.3 $\mu$m absorption band. We note however that the 11.3 $\mu$m band is most sensitive to the innermost regions of the shell since it is in absorption. Therefore the 11.3 $\mu$m band remains a robust indicator of the most recent mass-loss rate. But because of the dependence on the dust shell parameters the 33.6 $\mu$m band is not a good forsterite abundance indicator.

Another reason why the 33.6 $\mu$m band is not a practical indicator is that for very high mass-loss rates, typically bigger than $10^{-4}$ M$_{\odot}/yr$, all model curves converge and so become insensitive to the forsterite abundance.

\section{Measuring the forsterite abundance}
We apply our method to six spectra of high mass-loss rate AGB stars (see table \ref{tab_obs}) observed by ISO-SWS \citep{syl99}. The reduced data have been taken from \citet{sloan03}. We have used ISO-SWS spectra and not Spitzer spectra, since those are still in the pipe-line to be reduced and analyzed. Inspection of the spectra shows that the 9.7 $\mu$m amorphous silicate band is in absorption. In addition, a shoulder is seen in the observed spectra at 11.3 $\mu$m, which we attribute to forsterite. The latest detection of a very strong 11.3 $\mu$m band has been reported by \citet{speck08}. Their detection is the first detection where the 11.3 $\mu$m forsterite band does not appear as a shoulder on the 9.7 $\mu$m band of amorphous silicate, but as a separate absorption band. Given the presence of the 33.6 $\mu$m forsterite band in the ISO spectra of our sample of six stars \citep{kemper01}, we suggest that the 11.3 $\mu$m shoulder can indeed be attributed to forsterite as well.

We have also fitted the 9.7 $\mu$m band of amorphous silicate using the opacities discussed in section 2.1 and find good agreement (not shown). This suggests that the adopted opacities that were obtained by fitting the 9.7 $\mu$m band of Mira also apply to the higher mass-loss rate AGB stars. It also means that we can use these opacities in this section to measure the forsterite abundance.

We have measured the band strength of the 11.3 $\mu$m forsterite band as outlined in sect. 3.1. The errors calculated for the strength of the bands are determined using standard error propagation formula applied to the errors in the flux obtained from \citet{sloan03}. A 10\% error in the flux of the continuum is introduced to account for possible errors in the construction of the continuum. This is done for the two continua constructed in order to measure the strength of the 11.3 $\mu$m band. Mass-loss rates from the literature are listed in Table \ref{tab_obs}. No mass-loss rate was published for IRAS 17010$-$3840. The spectrum of IRAS 17010-3840 has its 9.7 $\mu$m band of amorphous silicate firmly in absorption (similar to the other sources), therefore its mass-loss rate has been assumed to lie between $5 \cdot 10^{-5}$ M$_{\odot}$ yr$^{-1}$ and $2 \cdot 10^{-4}$ M$_{\odot}$ yr$^{-1}$.

\begin{table}
\begin{minipage}[t]{\columnwidth}
\caption{A selection of AGB stars with their mass-loss rates and the results. It lists the strength of the 11.3 $\mu$m band together with the indicated forsterite (fo.) abundance using the 11.3 $\mu$m band. The sources have been selected using the catalog of \citet{sloan03}. The forsterite abundance obtained by \citet{kemper01} has also been included.}
\label{tab_obs} 
\centering 
\renewcommand{\footnoterule}{}  
\begin{tabular}{l l l l l l} 
\hline\hline 
Source		& 	\.{M} 			& Strength	& Fo. abun.		& Fo abun.		\\
		& 	(M$_{\odot}$ yr$^{-1}$)	& 11.3 $\mu$m	& 11.3 $\mu$m		& Kemper et 		\\
		&	$^{(}$\footnote{ \
			1: \citet{riechers05},\
			2: \citet{just96},\
			3: \citet{just92},\
			4: \citet{schut89},\
			5: \citet{groene94}}\
			$^{)}$ 			& band		& band			& al. (2001)			\\
\hline 
IRAS 17010-3840	&	?			& 0.14		&	$>$12\%	 	& -			\\
OH104.91+2.41	&	5.6 $10^{-5}$ (1)	& 0.03		&	$<$2\%		& $<$5\%		\\
OH26.5+0.6	&	16 $10^{-5}$ (2)	& 0.06		&	$\sim$4\%	& $<$5\%		\\
OH127.8+0.0	&	20 $10^{-5}$ (3)	& 0.04		&	$\sim$2\%	& $<$5\%		\\
OH30.1-0.7	&	15 $10^{-5}$ (4)	& 0.10		&	$\sim$9\%	& -			\\
OH32.8-0.3	&	16 $10^{-5}$ (5)	& 0.14		&	$>$12\%		& $<$5\%		\\
\hline 

\end{tabular}
\end{minipage}
\end{table}

\begin{figure}
	\resizebox{\hsize}{!}{\includegraphics{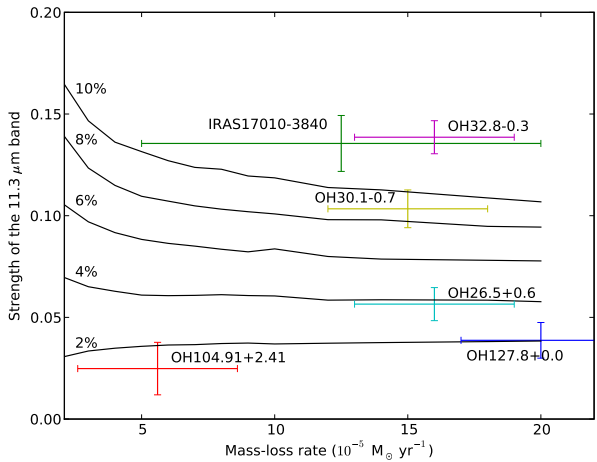}}
	\caption{The strength of the 11.3 $\mu$m band for six ISO-SWS sources (see table \ref{tab_obs}) together with the model results from fig. \ref{113_cryst_th}. All model parameters not mentioned in the graph are set to standard values (see table \ref{table:1}) }
	\label{113_OBS}
	\end{figure}

Figure \ref{113_OBS} shows the strength of the 11.3 $\mu$m band for the six sources. The forsterite abundances indicated by fig. \ref{113_OBS} are listed in table \ref{tab_obs}. The results show that both stars with a high and low forsterite abundance are found (below 2\% and up to $\sim$12\%). Because of our choice of thermal contact between the dust species these values are lower limits.

\citet{kemper01} have used the 33.6 $\mu$m band to measure the forsterite abundance in AGB stars. From their sample of six AGB stars (including OH104.91+2.41, OH127.8+0.0, OH26.5+0.6, OH32.8-0.3) they found a maximum forsterite abundance of 5\% by mass. The forsterite abundances we find for OH104.91+2.41, OH127.8+0.0 and OH26.5+0.6 do not exceed this maximum of 5\%. The sources OH32.8-0.3, OH30.1-0.7 and IRAS17010-3840 have significantly higher forsterite abundances.

\section{Conclusions}
We have shown that for optically thick dust shells the strength of the 11.3 $\mu$m band is a robust quantity to use as an indicator of the forsterite abundance of the current mass-loss period.

Applying the 33.6 $\mu$m band is practically impossible without knowledge of the mass-loss rate, the size of the dust shell and other parameters that influence the emission of the cold dust in the outskirts of the dust shell of the AGB star. Additionally, for certain mass-loss rates (depending on the dust shell parameters) the 33.6 $\mu$m band is also in transition from an emission to an absorption band, making the use of this band impossible.

As an example and first result we used the 11.3 $\mu$m band as an indicator of the forsterite abundance for six AGB stars. This showed that these objects can have low (below 2\%) but also very high forsterite abundances (12\% and possibly higher), meaning that AGB stars can have a forsterite abundance higher than 5\% by mass, which was the maximum found by \citet{kemper01}. It also means that the discrepancy between the amount of crystalline material in the ISM and that in AGB stars could even be higher. 

If the forsterite abundances found for the six sources in this study represent all AGB stars, the forsterite abundance in AGB stars alone would be higher than the crystalline abundance (upper limit of 2.2\%) found by \citet{kemperErr05} for the ISM. To find out if the abundances found for these stars are typical for AGB stars, more spectra have to be analyzed. The number of spectra are also not enough to test any correlations. But this paper opens the way to a more elaborate study similar to the one described in sect. 4 of this paper. Determining the amount of forsterite that is produced by high mass-loss rate AGB stars will help to test the suggested correlations and see if there is really a discrepancy between the produced abundance in AGB stars and the abundance found in the ISM.

\begin{acknowledgements} B.L. de Vries acknowledges support from the Fund for Scientific Research of
   Flanders (FWO) under grant number G.0470.07. 
\end{acknowledgements}

\bibliographystyle{aa}
\bibliography{references}


\end{document}